\documentclass[useAMS,usenatbib]{mn2e}
\usepackage{graphicx} 
\usepackage{color}

\def\be{\begin{equation}} 
\def\ee{\end{equation}}

\def\HI{\hbox{H~$\scriptstyle\rm I\ $}}

\def\gsim{\lower.5ex\hbox{\gtsima}} 
\def\lsim{\lower.5ex\hbox{\ltsima}} \def\gtsima{$\; \buildrel > \over 
\sim \;$} \def\ltsima{$\; \buildrel < \over \sim \;$} \def\prosima{$\; 
\buildrel \propto \over \sim \;$} \def\gsim{\lower.5ex\hbox{\gtsima}} 
\def\lsim{\lower.5ex\hbox{\ltsima}} 
\def\simgt{\lower.5ex\hbox{\gtsima}} 
\def\simlt{\lower.5ex\hbox{\ltsima}} 
\def\simpr{\lower.5ex\hbox{\prosima}} \def\la{\lsim} \def\ga{\gsim} 
  
 \def\gtsima{$\; \buildrel > \over \sim \;$} 
\def\ltsima{$\; \buildrel < \over \sim \;$} 
\def\gsim{\lower.5ex\hbox{\gtsima}} 
\def\lsim{\lower.5ex\hbox{\ltsima}} 
\def\simgt{\lower.5ex\hbox{\gtsima}} 
\def\simlt{\lower.5ex\hbox{\ltsima}} 
\def\simpr{\lower.5ex\hbox{\prosima}} 
\def\la{\lsim} 
\def\ga{\gsim}

\def\E3{{\cal E}_{\rm g}^{III}}

\def\Msun{\rm M_\odot}

\def\Msun{\rm M_\odot}

\def\M*{M_*}
\def\Z*{Z_*}
\def\L*{L_*}

\def\der{{\rm d}} 
\def\f{\frac}
\def\kev{\rm keV}
\def\mx{\,m_x} 
\def\K{\rm K}

\title[DCBH in CDM and WDM]{Warm dark matter constraints from high-$z$ Direct Collapse Black Holes using the JWST} 
\author[Dayal et al.]{Pratika Dayal$^{1,2}$\thanks{dayal@astro.rug.nl}, Tirthankar Roy Choudhury$^3$, Fabio Pacucci$^4$ \& Volker Bromm$^5$  \\ 
$^{{1}}$ Kapteyn Astronomical Institute, University of Groningen, P.O. Box 800, 9700 AV Groningen, The Netherlands\\
$^{2}$ Institute of Space Sciences \& Astronomy, University of Malta, Msida MSD 2080, Malta \\
$^3$ National Centre for Radio Astrophysics, Tata Institute of Fundamental Research, Pune 411007, India\\
$^4$ Department of Physics, Yale University, New Haven, CT 06511, USA \\
$^5$ Department of Astronomy, University of Texas, Austin, TX 78712, USA
}

\begin{document} 
 
\date{} 

\maketitle

\begin{abstract}
We use a semi-analytic model, {\it Delphi}, that jointly tracks the dark matter and baryonic assembly of high-redshift ($z \simeq 4-20$) galaxies to gain insight on the number density of Direct Collapse Black Hole (DCBH) hosts in three different cosmologies: the standard Cold Dark Matter (CDM) model and two Warm Dark Matter (WDM) models with particle masses of 3.5 and 1.5 keV. Using the Lyman-Werner (LW) luminosity of each galaxy from {\it Delphi} we use a clustering bias analysis to identify all, pristine halos with a virial temperature $T_{vir}\gsim10^4$ K that are irradiated by a LW background above a critical value as, DCBH hosts. In good agreement with previous studies, we find the DCBH number density rises from $\sim10^{-6.1}$ to $\sim 10^{-3.5}\, \mathrm{cMpc^{-3}}$ from $z\simeq 17.5$ to $8$ in the CDM model using a critical LW background value of $30 J_{21}$ (where $J_{21}= 10^{-21} \, {\rm erg\, s^{-1}\, Hz^{-1} \, cm^{-2} \, sr^{-1}}$). We find that a combination of delayed structure formation and an accelerated assembly of galaxies results in a later metal-enrichment and an accelerated build-up of the LW background in the 1.5 keV WDM model, resulting in DCBH hosts persisting down to much lower redshifts ($z \simeq 5$) as compared to CDM where DCBH hosts only exist down to $z \simeq 8$. We end by showing how the expected colours in three different bands of the Near Infrared Camera (NIRCam) onboard the forthcoming {\it James Webb Space Telescope} can be used to hunt for potential $z \simeq 5-9$ DCBHs, allowing hints on the WDM particle mass.

\end{abstract}

\begin{keywords}
Cosmology: dark ages; Galaxies: formation - high-z - haloes; Black Hole physics; methods - statistical
\end{keywords}

%#################################################################
\section{Introduction}
%#################################################################
The particle nature of Dark Matter (DM), that comprises $\sim$84\% of the total matter density of the Universe, remains one of the key outstanding questions in physical cosmology. The standard Lambda Cold Dark Matter ($\Lambda {\rm CDM}$) cosmological model consists of dark energy, dark matter and baryons with energy densities corresponding to $\Omega_\Lambda$ , $\Omega_m$ , $\Omega_b$ = $0.6911, 0.3089, 0.049$ \citep{planck2016a}. In this model, CDM is composed of massive, collissionless particles with very low thermal velocities leading to negligible free-streaming on structure-formation scales. Some of the most popular CDM candidates \citep[reviewed in][]{feng2010, peter2012} are: {\it (i) WIMPS:} with a mass between $10\, {\rm GeV}$-${\rm TeV}$ stable Weakly Interacting Massive Particles may be produced as a thermal relic of the Big Bang, with a relic density consistent with that required for DM, making them the preferred CDM candidate; {\it (ii) Axions:} with a mass $\sim 10^{-5} {\rm eV}$, although axions provide a natural solution to the strong charge-parity problem in particle physics, they require fine-tuning to match to the observed DM number density and {\it (iii) Gravitinos:} are a supersymmetric counterpart of the graviton whose mass could lie anywhere between the ${\rm eV}$ to ${\rm TeV}$ scales depending on how super-symmetry is broken, with $\gsim$ keV gravitinos being a possible CDM candidate.

Despite a lack of consensus regarding its exact particle nature, CDM has been extremely successful at explaining the evolution of the large scale structure of the Universe ranging from the Cosmic Microwave Background (CMB) to Lyman Alpha forest statistics to galaxy clustering to weak lensing \citep[e.g.][]{peebles1971, blumenthal1984, diemand2011, slosar2013, planck2016a}. However, CDM has an excess of small-scale power which manifests itself in a number of issues including the ``missing satellite problem" where simulations predict thousands of sub-halos of the Milky Way as opposed to the few tens detected observationally \citep[e.g.][]{klypin1999, moore1999b}, the ``core-cusp problem" where simulated CDM halos show central density profiles rising with radius, as $r^{-1}$ to $r^{-1.5}$, as compared to the constant density cores inferred observationally \citep[e.g.][]{navarro1997, moore1999} and the ``too big to fail" problem where observations show a lack of theoretically predicted bright, massive dwarf galaxies of the Milky Way with circular velocities $V_{vir} \gsim 20\, {\rm km\, s^{-1}}$ \citep[e.g.][]{boylan2012}. 

While baryonic feedback can alleviate some of these small-scale problems \citep[e.g.][]{maccio2012, governato2012, dicintio2014, governato2015}, other works have focused on revisiting the cold, collissionless nature of DM itself. Indeed, a number of authors have jointly tracked DM and baryons to study galaxy/structure formation and reionization in alternative models ranging from Warm Dark Matter \citep[WDM;][]{menci2012, lovell2014, lovell2017, dayal2015, magg2016, dayal2017} to fuzzy CDM \citep{du2017} to self-interacting DM \citep{rocha2013, vogelsberger2014} to decaying DM \citep{wang2014}. Of these, WDM is arguably the most popular scenario given that particle physics provides a physical motivation for light ($\sim \kev$) DM particles such as sterile neutrinos. Their case has been strengthened by observations of a $3.5\, \kev$ line from the Perseus galaxy cluster and in the Chandra deep fields that could potentially originate from a light ($\sim \kev$) sterile neutrino annihilating into photons \citep{bulbul2014, boyarsky2014, cappelluti2017}. Current constraints on the WDM particle mass range from $\mx \sim 2.1-5.3\, \kev$ using the Lyman Alpha forest \citep{viel2013, garzilli2015, irsic2017}, $\mx \gsim 1.6\, \kev$ using high-$z$ Gamma Ray Bursts \citep{desouza2013}, $\mx \gsim 1\kev$ from observations of dwarf spheroidal galaxies, stellar mass functions of $z =0-3.5$ galaxies and $z \sim 10$ Lyman Break Galaxies \citep[LBGs;][]{devega2010, kang2013, pacucci2013} and $\mx \gsim 1.3-2.1\, \kev$ using a variety of techniques to model $z \gsim 6$ LBGs \citep[e.g.][]{schultz2014, lapi2015, menci2016a, menci2016b, corasaniti2017, dayal2017}. Finally, a number of probes, including low-redshift velocity functions \citep[see e.g.][]{schneider2017} and the abundance of satellite galaxies \citep{poli2011, kennedy2014, horiuchi2016} yield $m_x>1.5\kev$ independent of baryonic processes \citep[for a discussion see e.g.][]{schneider2016}.

In this work we study the number density of {\it Direct Collapse Black Holes (DCBHs)} at $z \simeq 4-20$ in three DM models: CDM and WDM with particle masses of 3.5 and 1.5 keV. Our aim is to understand if DCBHs can possibly be used to constrain the WDM particle mass using the unique observational capabilities of the forthcoming {\it James Webb Space Telescope} (JWST). First postulated as massive ($10^{5-6}\Msun$) black hole seeds, to explain the presence of super-massive black holes (SMBH) at early cosmic epochs \citep[e.g.][]{loeb1994, bromm2003}, DCBH formation scenarios have been continually refined and developed over the past years \citep[e.g][]{begelman2006, regan2009, Shang_2010, johnson2012, latif2013, agarwal2014, dijkstra2014, ferrara2014, habouzit2016}. This field has been lent impetus by the possible, albeit highly debated \citep{sobral2015, bowler2017}, detection of a DCBH \citep{agarwal2016, hartwig2016, smith2016, Pacucci_2017} in the Lyman Alpha Emitting ``CR7" galaxy at $z \sim 7$. The current understanding built from these works requires the following conditions be met for a halo to host a DCBH: (i) the halo should have reached the atomic cooling threshold with a virial temperature $T_{vir}\gsim 10^4 {\rm K}$ so that gas can cool isothermally; (ii) the halo should be metal-free to prevent gas fragmentation; and (iii) the halo should be exposed to a high enough ``critical" Lyman-Werner (LW) background ($J_{crit} = \alpha J_{21}$) where $\alpha>1$ is a free parameter and $J_{21}$ is the LW background expressed in units of $10^{-21} {\rm erg\, s^{-1}\, Hz^{-1} \, cm^{-2} \, sr^{-1}}$ (see e.g. \citealt{Sugimura_2014}). 
%Moreover, a tentative selection of DCBH candidates in the CANDELS/GOODS-S field was already done in \cite{pacucci2016b}. 

Our aim of combining the fields of DCBH and WDM research is based on the following argument: while the delay in structure formation in WDM cosmologies would result in a later metal-enrichment, the faster stellar mass assembly with decreasing redshift \citep[e.g. Fig. 3;][]{dayal2015} would lead to a correspondingly steeper build-up of the LW background. Together, these effects might collude to create ideal conditions for DCBH formation for a longer cosmic epoch. The aim of this paper is two-fold: firstly, we study the conditions for DCBH formation as a function of the underlying DM cosmology. Secondly, we investigate if the existence of DCBHs down to lower-$z$ in WDM cosmologies could be tested by forthcoming facilities, including the JWST, yielding constraints on the WDM particle mass, complementary to the Lyman Alpha forest, at an epoch inaccessible by other means.

We use a semi-analytic model to jointly track the growth of DM halos and their baryonic component in both CDM and WDM cosmologies as explained in Sec. \ref{theo_model} that follows. We obtain estimates of metal-free halos from our semi-analytic model as explained in Sec. \ref{met}. We use stellar-mass assembly histories from the model to quantify the LW background and its fluctuations as a function of both redshift and intrinsic galaxy properties as explained in Sec. \ref{sec_lwbg}. At the end of these calculations, we identify all DCBH hosts from $z \sim 4-20$ in the different DM models explored. Finally, we propose methods of testing the DM cosmology using the observed abundance of DCBHs and their colours in three different JWST bands in Sec. \ref{results}. In this work we solely focus on thermally-decoupled relativistic particles as WDM and investigate two scenarios with particle masses of $\mx=1.5$ and $3.5\kev$; these correspond to sterile neutrino masses of $m_\nu = 7.6$ and $23.4\, \kev$, respectively \citep{viel2005}.

%#################################################################
\section{Theoretical model}
\label{theo_model}
%#################################################################
This work is based on using the code {\it Delphi} ({\bf D}ark Matter and the {\bf e}mergence of ga{\bf l}axies in the e{\bf p}oc{\bf h} of re{\bf i}onization), introduced in \citet{dayal2014, dayal2015, dayal2017}. In brief, {\it Delphi} uses a binary merger tree approach to jointly track the build-up of DM halos and their baryonic component (both gas and stellar mass) through cosmic time. We start by building merger trees for 800 $z=4$ galaxies, uniformly distributed in the halo mass range of $\log(M_h/ \Msun)=8-13.5$, up to $z=20$. Each $z=4$ halo is assigned a co-moving number density by matching to the $\der n / \der M_h$ value of the $z=4$ Sheth-Tormen halo mass function (HMF) and every progenitor halo is assigned the number density of its $z=4$ parent halo; we have confirmed that the resulting HMFs are compatible with the Sheth-Tormen HMF at all $z$. The number density of halos at a given redshift is then simply the integral of the HMF over the relevant halo mass range such that $N_h(z) = \int \der M_h~\f{\der n}{\der M_h}$.

In what follows, we refer to the very first progenitors of a specific halo, that mark the start of its assembly, as the {\it starting leaves}. We start by assigning each such starting leaf an initial gas mass that scales with the halo mass according to the cosmological ratio such that $M_g = (\Omega_b/\Omega_m) M_h$; a fraction of this gas mass is converted into stars depending on the effective star formation efficiency of the host halo. The effective star formation efficiency, $f_*^{eff}$, for any halo is calculated as the minimum between the efficiency that produces enough type II supernova (SNII) energy to eject the rest of the gas, $f_*^{ej}$, and an upper maximum threshold, $f_*$, so that $f_*^{eff} = min[f_*^{ej}, f_*]$. The instantaneous stellar mass produced at any $z$ is then calculated as $M_*(z) = f_*^{eff} M_g(z)$. The final gas mass, at the end of that $z$-step, including the effects of star formation and supernova feedback, is then given by $M_{gf}(z) = [M_g(z)-M_*(z)] [1-(f_*^{eff}/f_*^{ej})]$. At each $z$-step we also account for DM that is smoothly accreted from the inter-galactic medium (IGM); we make the reasonable assumption that such smooth-accretion of DM mass is accompanied by accretion of a cosmological fraction ($\Omega_b/\Omega_m$) of gas mass from the IGM. We use a  using a Salpeter initial mass function \citep[IMF;][]{salpeter1955} between $0.1-100\Msun$ throughout this work. Implementing this physical prescription, we have shown our model to be in excellent agreement with all available observables for high-$z$ ($z \gsim 5$) galaxies, including the evolving Ultra-violet luminosity function (UV LF), the stellar mass function, the mass-to-light ratios and the $z$-evolution of the stellar mass and UV luminosity densities, for both cold and warm dark matter cosmologies. We note that the model only uses two mass- and $z$-independent free parameters: to match to observations we require roughly 10\% of the SNII energy coupling to the gas and a maximum (instantaneous) star formation efficiency of $f_* = 3\%$.

%#################################################################
\subsection{Identifying pristine halos }
\label{met}
%#################################################################
We start by identifying all halos with $T_{vir}\gsim10^4 \K$.  We then make the reasonable assumption that all the starting leaves of any halo are metal-free by virtue of never having accreted metal-enriched gas. In order to study the external metal-enrichment driven by SNII outflows from nearby galaxies, we calculate the total radius of metal-enrichment, $R_{m}$, for each galaxy, with stellar mass $M_*$ at a redshift $z$, as \citep[e.g.][]{dijkstra2014}:
\begin{equation}
R_m (M_*,z) = \bigg(\frac{E_{51}\nu M_* t_6^2}{m_p n}\bigg)^{1/5} \, {\rm kpc}, 
\end{equation}
where $E_{51}=10^{51} \,{\rm erg}$ is the SNII explosion energy, $\nu = [134\Msun]^{-1}$ is the SNII rate for the assumed IMF and $m_p$ is the proton mass. Further, $n$ is the number density of gas in which the SNII goes off for which we assume a value of $60 \rho_c(z)$ where $\rho_c(z)$ is the critical density at $z$ and we use an age of $t_6=20 \, {\rm Myr}$ given our merger tree time-steps of 20 Myrs. Comparing this value to the virial radius, we find that $R_m(M_*,z) = 0.5-2 R_{vir} (M_*,z)$ for CDM and has a slightly narrower range of $R_m(M_*,z) = 0.5-1.5 R_{vir} (M_*,z)$ for 1.5 keV WDM, justifying our assumption of ignoring externally-driven metal enrichment. However, as noted in \citet{dijkstra2014}, although initiated in the dense interstellar-medium, the SNII blast wave will spend most of its time a much lower density IGM - the value of $R_m$ calculated here is therefore a lower limit on the metal-enrichment radius. This naturally implies that a fraction of our starting leaves could have been enriched by accretion of pre-enriched IGM gas - our values of ``pristine" halos should therefore be treated as an upper-limit on the total number density. Further, many of these leaves could have had mini-halo progenitors of their own using a higher resolution merger-tree. While realistic estimates of metal-enrichment would require at least N-body, if not hydrodynamic, simulations, beyond the scope of this work, our simplistic assumption on the upper-limit of DCBH hosts is justified, and compensated for, by exploring results for $\alpha$ varying over an order of magnitude, such that $J_{crit} = 30-300 J_{21}$. We note that the number density of pristine halos could change by a factor of 10 only if 90\% of them were metal-enriched; this is highly unlikely given the small $R_m$ values calculated above. Further, mini-halo progenitors of starting halos would not be viable DCBH hosts given that they would not be irradiated by the required LW intensity of $J_{crit} \gsim 30 J_{21}$. 

%#################################################################
\subsection{The Lyman-Werner background and its fluctuations}
\label{sec_lwbg}
%#################################################################

We use the stellar population synthesis code {\it Starburst99} \citep{leitherer1999} to calculate the LW ($11.2-13.6$ eV) luminosity of each galaxy based on its entire star formation history. We use this to calculate the mean LW emissivity, $\epsilon_{\rm LW}(z)$, at a given redshift by integrating over all galaxies present at that $z$. The mean LW background intensity can then be calculated as
\be
\bar{J}_{\rm LW}(z) = \frac{(1 + z)^2}{4 \pi} \int_z^{\infty} \frac{c~\der z'}{H(z')}~\epsilon_{\rm LW}(z)~f_{\rm mod}(r_{\rm com}(z,z')),
\label{lwbg}
\ee
where $H(z')$ is the Hubble parameter at redshift $z'$, $c$ denotes the speed of light, $r_{\rm com}(z,z')$ is the radial comoving distance between the two redshifts $z$ and $z'$, and $f_{\rm mod}(r)$ is the LW flux profile as a function of the comoving distance from the source. We use  the fitting form of $f_{\rm mod}(r)$ given by \cite{ahn2009} based on their radiative transfer simulations.

Unfortunately the mean background cannot be used for estimating the fraction of galaxies affected by LW radiation. This is because of the fluctuations in the background which may lead to small patches, most likely around the galaxies, where the intensity is above the critical threshold value $J_{crit}$ even when the background intensity is smaller. It is possible to calculate the probability that a collapsing halo forms within a region where the LW background is above the critical value using an approach based on the halo model \citep{holz2012}. 

The first step is to assign to each galaxy a threshold radius $r_{\rm thres}$ within which the LW background from that galaxy exceeds $J_{crit}$\footnote{The LW background intensity profile for a galaxy can be calculated as \citep{ahn2009,dijkstra2014}
\be
J_{\rm LW}(r) = (1 + z)^2~\f{L_{\rm LW}}{16 \pi^2 r^2}~f_{\rm mod}(r),
\ee
where $r$ is the comoving distance from the source. The threshold radius is given by the equation $J_{\rm LW}(r_{\rm thres}) = J_{crit}$. 
}.
In case the sources are very rare (i.e., when we can ignore the overlap of LW threshold regions around individual sources), the volume fraction of regions with $J_{\rm LW} > J_{crit}$ is given by
\be
Q'_{\rm thres}(z) = \int \der M_h ~ V_{\rm thres}~ \f{\der n}{\der M_h},
\ee
where $V_{\rm thres} \equiv 4 \pi r_{\rm thres}^3 / 3$ is the volume of the LW threshold region around the halo.

\begin{figure}
\center{\includegraphics[scale=0.48]{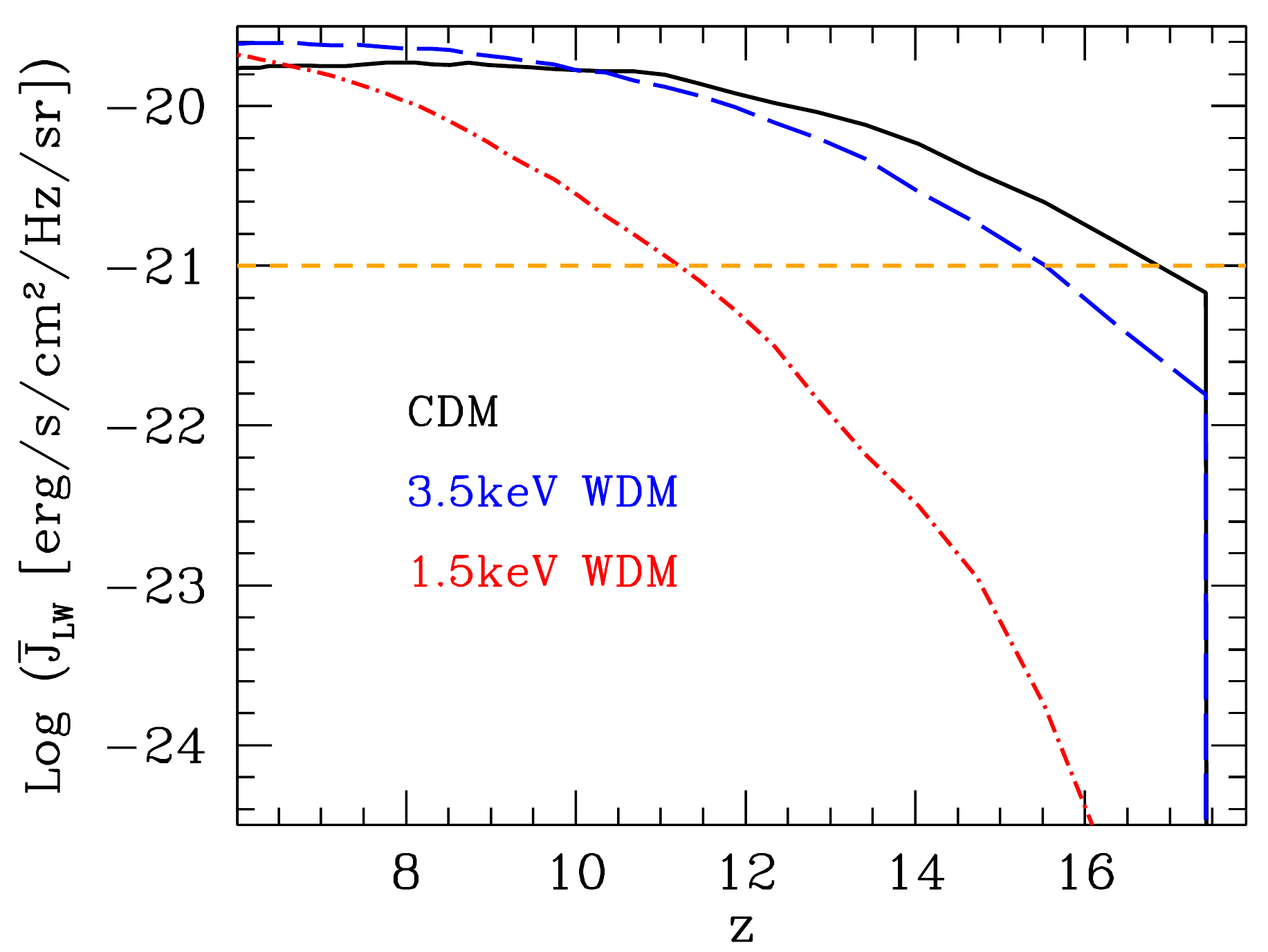}}
\caption{The mean LW background as a function of redshift for the three different DM cosmologies studied in this work: CDM (solid black line), 3.5 keV WDM (dashed blue line) and 1.5 keV WDM (dot-dashed red line). We use standard units wherein $J_{21} = 10^{-21} {\rm erg\, s^{-1}\, Hz^{-1} \, cm^{-2} \, sr^{-1}}$ as shown by the dashed horizontal line. As seen, although the background starts building up later in the 1.5 keV WDM scenario, its subsequent evolution is more accelerated as compared to the other two models (see Sec. \ref{res_lwbg}). }
\label{jlw} 
\end{figure}

\begin{figure*}
\center{\includegraphics[scale=1.01]{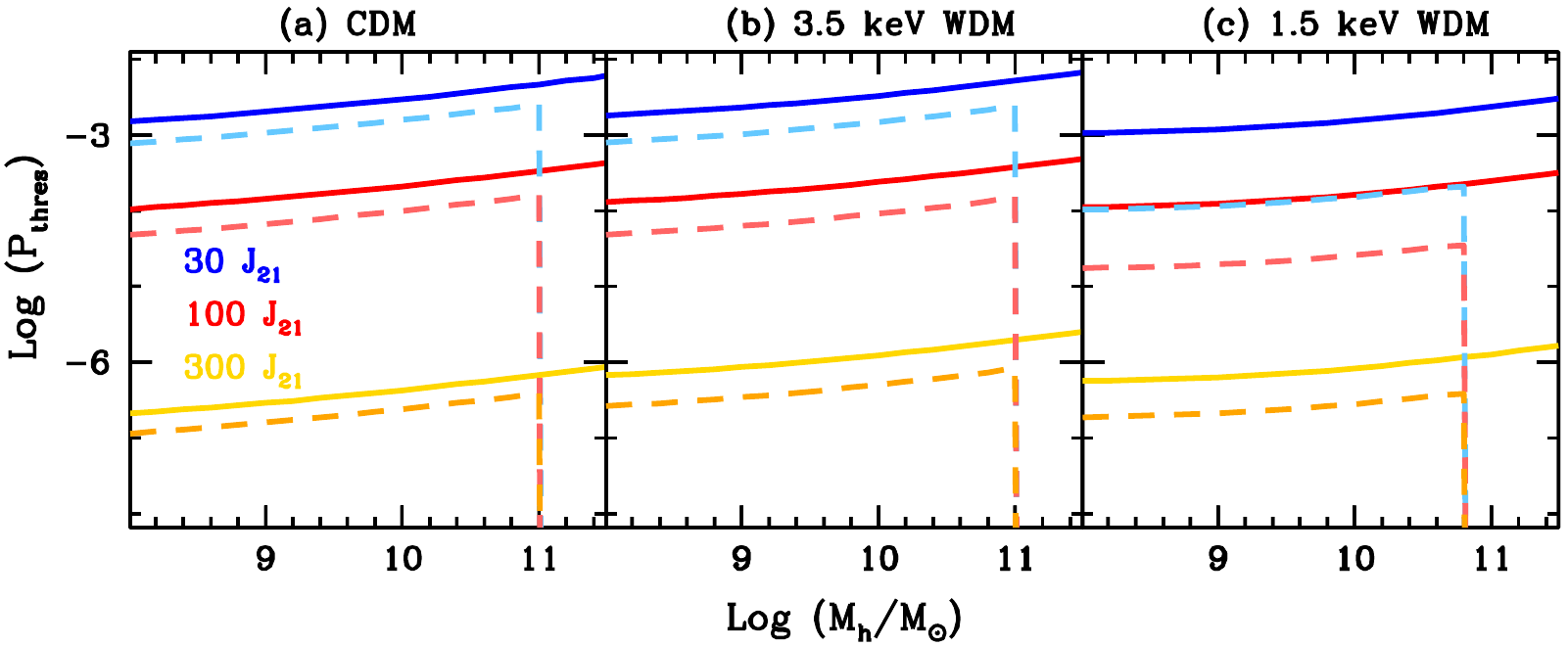}}
\caption{The (log) probability of a given halo mass being irradiated by a LW background exceeding a chosen critical value for the three DM models considered in this work, as marked above each panel. The solid blue, red and yellow lines show results at $z\sim 9$ for critical LW values corresponding to $30, 100$ and $300J_{21}$, respectively; the dashed lines show the corresponding results at $z \sim 12$. }
\label{prob_jcrit} 
\end{figure*}

In order to calculate the probability that a new halo forms in region with $J_{\rm LW} > J_{crit}$, we need to account for the fact that haloes are preferentially biased and hence are more probable to lie closer to other haloes. Let $\bar{r}_{\rm thres} \equiv \left[3 \bar{V}_{\rm thres} / (4 \pi) \right]^{1/3}$ be the average value of the threshold radius, where the averaging is done over all haloes. Then the excess probability that two haloes exist within a comoving distance $\bar{r}_{\rm thres}$ is given by $b_1~b_2~\xi_{\rm DM}(\bar{r}_{\rm thres})$, where $b_1, b_2$ are the bias of the two haloes and $\xi_{\rm DM}$ is the correlation function of the dark matter density field. For simplicity, we use linear theory to calculate the halo bias \citep{sheth-tormen1999} and $\xi_{\rm DM}$ for all the DM models considered in the paper. Hence the probability that a halo of mass $M_h$ lies within a distance $\bar{r}_{\rm thres}$ of an existing halo is
\be
Q_{\rm thres}(M_h, z) = Q'_{\rm thres}(z) [1 + b(M_h, z) \bar{b}_{\rm LW}(z)~\xi_{\rm DM}(\bar{r}_{\rm thres}, z)],
\ee
where $b(M_h,z)$ is the halo bias and
\be
\bar{b}_{\rm LW}(z) = \f{\int \der M'_h~(\der n / \der M'_h)~b(M'_h, z)~V_{\rm thres}}{\int \der M'_h~(\der n / \der M'_h)~V_{\rm thres}}
\ee
is the mean bias of LW threshold regions around the existing galaxies. If the LW producing galaxies are assumed to be randomly distributed, the probability is given by $P_{\rm thres} = 1 - \exp(- Q_{\rm thres})$.

Our calculation of the LW threshold probability does not account properly for clustering of the sources and resulting overlap of the threshold volumes, which require more complex methods such as high dynamic range $N$-body simulations or Monte-Carlo realisations \citep{dijkstra2008,dijkstra2014}. Also, our treatment of the halo bias and correlation function becomes less accurate at small scales and thus will be important for higher values of $J_{crit}$ which lead to smaller values of $\bar{r}_{\rm thres}$. A possible improvement can be achieved by using more sophisticated characterisation of the bias \citep{dijkstra2008,jose2017}. As noted above, we compensate for these uncertainties by exploring a large range (a factor of 10) in $\alpha$ - indeed, even  scale-dependent clustering would not be expected to affect our estimates of $P_{thres}$ by one order of magnitude.

With this model, the total number density of possible DCBH hosts, $N_h$, at any $z$ is calculated as
\begin{equation}
N_h(z) = \int dM_h \frac{dn}{dM_h} \times P_{T4} \times P_{leaf} \times P_{\rm thres}, 
\label{nd_dcbh}
\end{equation}
where $P_{T4}$ and $P_{leaf}$ are the binary probabilities of a halo having $T_{vir}>10^4 \K$ and being a starting leaf, respectively. 

%calculate the number density of all halos, all halos with $T_{vir}>10^4 K$, all leaves  and all leaves with $T_{vir}>10^4 K$, as shown in Fig. \ref{nd_dm}. We make the assumption that the leaves with $T_{vir}>10^4K$ fulfil two of the three conditions necessary to host DCBH: (i) their masses are large enough to accrete a sufficient amount of gas and (ii) they are, by definition, metal-free, never having undergone star-formation. We therefore mark these as ``potential DCBH hosts" which are not allowed to form stars. We have checked this has no bearing on our results, given their low number-densities compared to all halos at a given $z$. 

%#################################################################
\section{Results}
\label{results}
%#################################################################
We now discuss the results obtained from our model. We start by discussing the build-up of the LW background and its implications for DCBH hosts before discussing the evolution of their number density through cosmic time in the three DM models considered in this work.

%************
\subsection{Redshift-evolution of the LW background and implications for DCBH hosts}
\label{res_lwbg}
%************
We start by studying the redshift-evolution of $\bar J_{\rm LW}$ (see Eqn. \ref{lwbg}), in Fig. \ref{jlw}, in all the three DM models considered in this work. As seen, galaxy formation and, the build-up of the LW background starts as early as $z \sim 17.5$ in CDM; the $3.5\, \kev$ WDM particle is massive enough for its LW background to be essentially indistinguishable from CDM at all $z$. On the other hand, a delay in structure formation in the 1.5 keV WDM model \citep[see e.g. Fig. 9][]{dayal2017} results in a later build-up of the LW background with, the lack of low mass halos resulting in, a lower amplitude as compared to CDM at $z \gsim 7$. Quantitatively, $\bar J_{\rm LW}$ is lower by an order of magnitude in the $1.5\, \kev$ WDM model when compared to CDM at $z \simeq 10$, increasing to a difference of about three orders of magnitude at $z \sim 15$. Further, while galaxies in CDM and 3.5 keV WDM models reach the value of $J_{21}$ at $z \sim 15.5-16.5$, galaxies in the 1.5 keV WDM model reach this value, after a delay of about 150 Myrs, at $z \simeq 11$ - this corresponds to a delay of roughly 30\% of the age of the Universe at $z \simeq 11$. We also note that the lack of low-mass, feedback-limited, progenitors results in a steeper slope for the LW background in the $1.5\, \kev$ WDM model which has a higher amplitude at $z \lsim 7$, compared to CDM, analogous to the steeper slope of the stellar mass density \citep[see Fig. 6;][]{dayal2015}.

We then quantify the probability of galaxies of a given halo mass being irradiated by LW background with a critical $J_{crit} = \alpha J_{21}$ and explore $\alpha = 30,100, 300$ so as to bracket the range studied in earlier works \citep[e.g.][]{agarwal2012, dijkstra2014, habouzit2016}. Given that the LW background produced by a galaxy dominates within the threshold radius, it might be expected that galaxies with a larger halo mass will have a higher probability of attaining $P_{\rm thres}$. This is indeed the case as seen from Fig. \ref{prob_jcrit} where, independent of the DM cosmology and redshift, $P_{\rm thres}$ increases by a factor of 3 as the halo mass increases from $10^{8}$ to $\simeq 10^{11-11.5}\Msun$. As expected, the value of $P_{\rm thres}$ decreases with increasing $\alpha$ - quantitatively we find that, independent of cosmology, $P_{\rm thres}$ decreases by about one order of magnitude as $\alpha$ increases from 30 to 100 with a steep drop of about four orders of magnitude as $\alpha$ further increases to $300$ showing the severe lack of galaxies being irradiated by such intense LW fields. Although the probability values only go up to $M_h \simeq 10^{11}\Msun$ at $z \simeq 12$, since higher mass galaxies not have yet assembled at such early epochs, the qualitative results remain unchanged up to this redshift. Finally, as a result of the lower value of $\bar J_{LW}$, $P_{\rm thres}$ values are slightly lower for all $\alpha$ values in the 1.5 keV WDM model. However, this difference is too small to have any appreciable effect on the results that follow.

\begin{figure*}
\center{\includegraphics[scale=1.01]{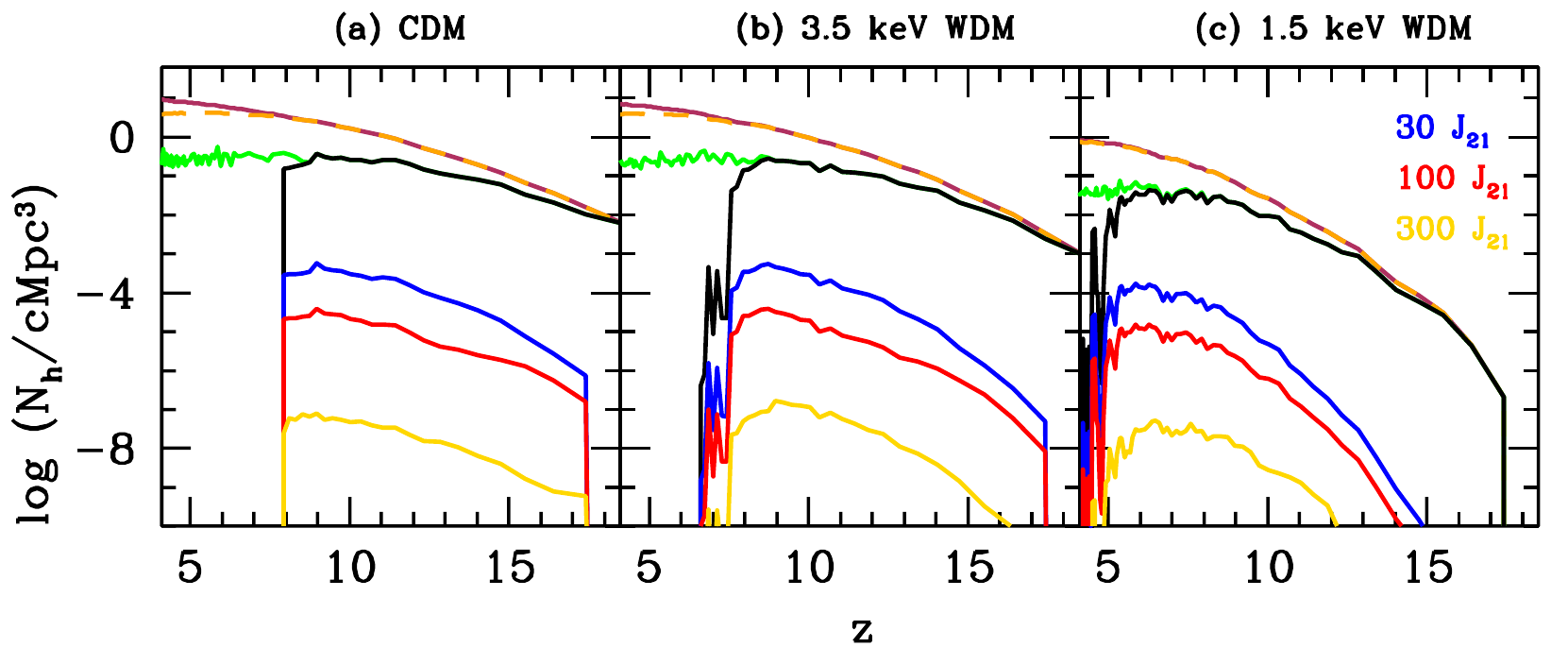}}
\caption{Number density of DM halos and DCBH hosts as a function of z, for CDM, 3.5 keV and 1.5 keV WDM, as marked above each panel. In each panel, the solid maroon line shows the number density for all halos at that $z$, the dashed orange line shows the number density of halos with $T_{vir}\gsim10^4 \K$ and the green line shows the number density of ``starting leaves" that are, by definition, assumed to be metal-free. The black line shows the number density of metal-free leaves with $T_{vir}\gsim10^4 \K$ - this represents the upper limit on the number density of DCBH hosts. Finally, the solid blue, red and yellow lines show the number density of DCBH hosts irradiated by a minimum LW background corresponding to $30, 100$ and $300J_{21}$, respectively.}
\label{nd_hosts} 
\end{figure*}

%************
\subsection{Redshift-evolution of DCBH host number densities}
%************
We now discuss the number density of DCBH hosts, obtained from the calculations above, shown in Fig. \ref{nd_hosts}. We start by looking at the number density of all halos from the merger-tree as a function of $z$ which rise by about three orders of magnitude from $\sim 10^{-2}$ to $\sim 10 \, \mathrm{cMpc^{-3}}$ from $z \simeq 17.5$ to $z \simeq 4$ in CDM; given the lack of small-scale structure at all epochs, the corresponding values are lower in 1.5 keV WDM, rising from $\sim 10^{-6.5}$ to $\sim 1 \, \mathrm{cMpc^{-3}}$ from $z \simeq 17.5$ to $z \simeq 4$. Effectively all these halos have $T_{vir}\gsim 10^4 \, \K$, given the minimum halo mass of $10^8 \, \Msun$ used in the merger-tree, so that cutting-off all halos below this virial temperature has no sensible effect, as shown in the same figure. However, the number density of the starting leaves decreases with decreasing $z$: while starting leaves comprise (almost) all of the halos at $z \simeq 17.5$, this value drops sharply such that they comprise only about 3\% of the total number density by $z \simeq 4$, independent of the DM model. 

We then look at the number density of starting leaves with $T_{vir}\gsim 10^4 \K$ which effectively represents the upper limit on the number density of DCBH hosts (solid black line in Fig. \ref{nd_hosts}). Our results show an interesting dependence on the underlying DM model: while DCBH hosts can persist down to redshifts as low as $z\simeq 4$ in 1.5 keV WDM given the delay in structure formation, there is a sharp drop-off in the number density of DCBH hosts as early as $z \simeq 8$ in CDM; as expected, the 3.5 keV WDM model lies between the range bracketed by these two models, with DCBH hosts existing down to $z \simeq 6.5$. The sharp cut-off, at $z \simeq 8$, seen in the CDM model represents a physical limit: $T_{vir} = 10^4 \K$ corresponds to a halo mass of $M_h \sim 10^8 \Msun$ at $z = 8$. Given that CDM can collapse on all scales, all halos with $M_h \gsim 10^8 \Msun$, corresponding to $T_{vir} = 10^4 \K$, at lower-$z$ can be broken up into progenitors, i.e. by definition, they can not be the starting leaves. On the other hand, 1.5 keV WDM has a much larger collapse scale of about $10^9 \Msun$ - as a result, many $T_{vir} \gsim 10^4 \K$ halos remain un-fragmented (i.e. can be the starting leaves) and can act as DCBH hosts down to $z \simeq 4$. 

As shown in Eqn. \ref{nd_dcbh}, however, the {\it actual} number density of DCBH hosts sensitively depends on $P_{\rm thres}$. Accounting for $P_{\rm thres}$ results in the number density dropping by a factor of $\sim 10^3$ to as much as $10^{6.5}$ as $\alpha$ increases from a value of 30 to 300. Quantifying the values for CDM, while the number density of DCBH hosts rises from $\sim10^{-6.1}$ to $\sim 10^{-3.5}\, \mathrm{cMpc^{-3}}$ from $z\simeq 17.5$ to $8$ for $\alpha =30$, the corresponding number densities rise from $\sim 10^{-9.2}$ ($z \simeq 17.5$) to $\sim 10^{-7.5} \, \mathrm{cMpc^{-3}}$ ($z\simeq 8$) for $\alpha = 300$. Interestingly, although the final number densities of DCBH hosts exposed to a given LW background intensity are quite comparable in all the three DM models considered here (to within an order of magnitude) for $z \simeq 8-10$, the persistence of such hosts down to later cosmic epochs is a key detectable signature of light ($\sim 1.5$) keV WDM models, as discussed in the Sec. \ref{det} that follows. 
 
%#################################################################
\subsection{DCBH detectability}
\label{det}
%#################################################################

 \begin{figure*}
\center{\includegraphics[scale=0.99]{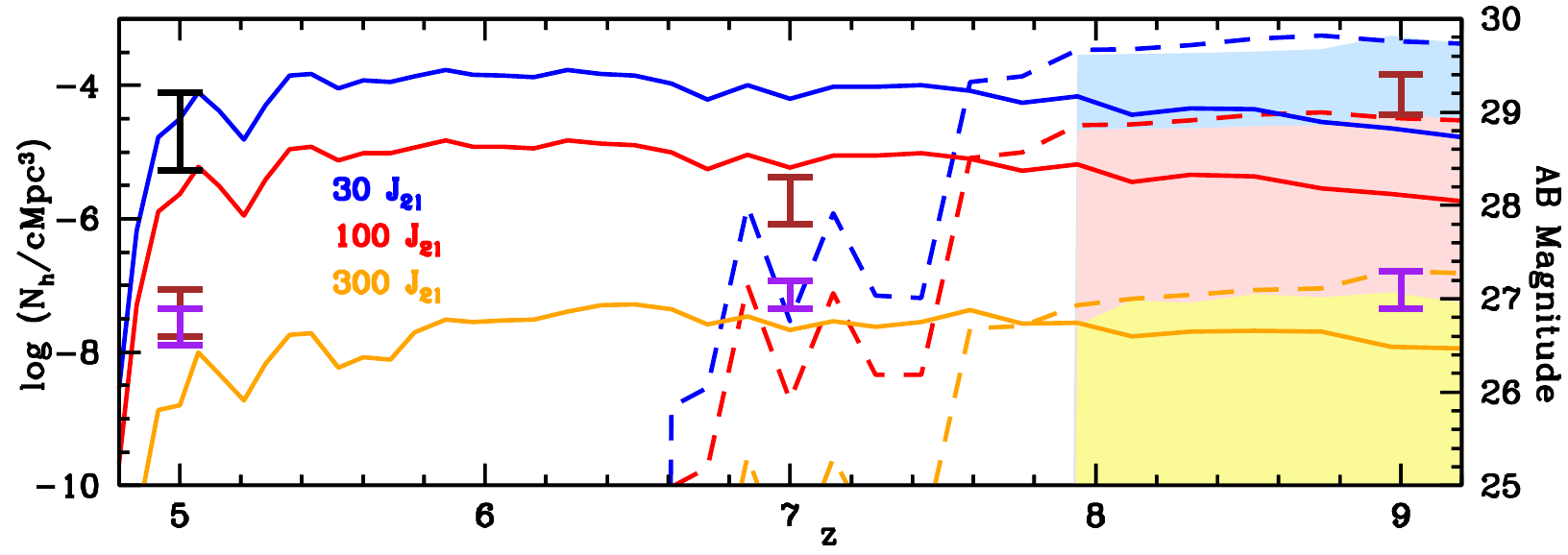}}
\caption{Number density of DCBH hosts as a function of z, for CDM (shaded region), 3.5 keV (dashed lines) and 1.5 keV WDM (solid lines) for three values of the critical LW background corresponding to $\alpha=30$ (blue lines), $\alpha=100$ (red lines) and $\alpha = 300$ (yellow lines), respectively. Potential DCBH candidates can be identified by their JWST colours shown on the right axes: here, symbols show the AB magnitude range expected for a DCBH mass between $10^{5-6}\Msun$ in the F070 band (black), F090 band (brown) and F115 band (purple) using the sensitivity threshold for a 10 ks observation. With these sensitivity limits, DCBHs would be undetectable in the F070W band at $z \gsim 7$ due to the attenuation from an increasingly neutral IGM. } 
\label{det_fig} 
\end{figure*}

Given the later persistence of DCBH hosts in 1.5 keV WDM models, we now discuss their specific observational imprints for which we focus on the redshift range $z \simeq 5-9$. A number of works \citep{pacucci2016b, natarajan2017, Volonteri_2017} have developed intricate photometric selection criteria for DCBHs, especially in context of the JWST. In this work, we use the models developed in \citet{pacucci2015} which, in brief, use the radiation-hydrodynamic code GEMS \citep{pacucci2015b} that models the accretion process onto the DCBH seed. We assume the standard $\alpha$-accretion disk scenario \citep{shakura1973} where the disk is optically-thick and geometrically thin. These results are coupled with the spectral synthesis code {CLOUDY \citep{ferland2013} to compute the complete spectrum emerging from the host halo of the DCBH. In order to obtain the ``observed" spectrum, we account for IGM attenuation using the neutral hydrogen fractions from {\it Delphi} that simultaneously match the observed CMB electron-scattering optical depth and emissivity constraints\footnote{We calculate the average neutral hydrogen fraction $\chi_{HI}$ from the volume filling fraction of ionized hydrogen ($Q_{II}$) shown in Fig. 4 of \citet{dayal2017} as 
\begin{equation}
\chi_{HI} = Q_{II} \times 0 + (1-Q_{II}) \times 1 = (1-Q_{II}), 
\end{equation}
making the simple assumption that while all the gas in ionized regions has been completely ionized, it remains completely neutral in non-ionized patches. To quantify, $Q_{II} \simeq (0.5,0.8,1)$ at $z \simeq (5,7,9)$ from our model.}. These theoretical spectra are then convolved with the complete filter transmission functions for the Near Infrared Camera (NIRCam) instrument on the JWST \footnote{https://jwst-docs.stsci.edu/display/JTI/NIRCam+Filters}- given our focus on $z \simeq 5-9$, we focus on the photometric colours in the F070W, F090W and F115W filters that bracket rest-frame wavelengths blue-ward of the Lyman Limit up to the UV. 

\begin{table}
\begin{center}
\caption{As a function of redshift (column 1), we show the AB magnitudes expected in three JWST bands (columns 3-5) for a DCBH of mass $10^5$ and $10^6\Msun$ (column 2) using the sensitivity threshold for a 10 ks observation. ``U" signifies DCBHs that would be undetectable with the JWST using this sensitivity threshold.}
\begin{tabular}{|c|c|c|c|c}
\hline
Redshift & DCBH mass $[\Msun]$ & F070W & F090W & F115W \\
\hline
$5$ & $10^5$ & $29.2$ & $27.1$ & $26.9$ \\
$ $ & $10^6$ & $28.3$ & $26.6$ & $26.5$ \\
$7$ & $10^5$ & $U$ & $28.2$ & $27.2$ \\
$ $ & $10^6$ & $U$ & $27.7$ & $26.9$ \\
$9$ & $10^5$ & $U$ & $29.4$ & $27.3$ \\
$ $ & $10^6$ & $U$ & $28.9$ & $26.8$ \\
\hline
\end{tabular}
\label{table1}
\end{center}
\end{table}

The results of these calculations are shown in Fig. \ref{det_fig} that reiterates the similarities between the number densities of DCBH hosts in all three DM models (to within an order of magnitude) at $z \gsim 8-10$ and the unambiguous persistence of DCBH hosts down to $z \sim 5$ in the 1.5 keV WDM scenario. Given that the exact number density depends on the value of $\alpha$, while volumes of the order of $(45\, \mathrm{cMpc})^3$ would suffice for $\alpha \lsim 100$, extreme cases of $\alpha =300$ would require surveying much larger volumes of about $(640\, \mathrm{cMpc})^3$ to hunt for DCBH hosts at $z \simeq 8-10$ in all three DM models. 

We then use the imprints of IGM attenuation to isolate detectable signatures of such DCBHs: the increasing neutrality of the IGM at $z \gsim 7$ results in a sharp break in flux blue-ward of the Lyman Alpha (Ly$\alpha$) line, at $1216 \, \mathrm{\AA}$ in the galaxy-rest frame; this break disappears at $z \lsim 6$ when the \HI in the IGM is mostly ionized. As shown in Table \ref{table1}, with the sensitivity threshold for a 10 ks JWST observation, we predict DCBHs at all $z \simeq 5-9$ to be detectable with AB magnitudes $\sim 26.5-29.5$ in the F090W and F115W bands that lie red-ward of the Ly$\alpha$. As expected, DCBH hosts are undetectable in the F070W filter at $z\gsim 7$, i.e. before the end of the reionization, independent of their mass. We therefore predict that while DCBHs could be detectable in all three JWST bands considered for $z \simeq 5-9$ for 1.5 keV WDM, DCBHs would not be detectable in the F070W band at $z \gsim 7$ in CDM and 3.5 keV WDM. Such colors therefore offer a testable means of hunting for DCBH hosts with further spectroscopy, for example with the JWST, required to unambiguously pin down their true nature. 

%#################################################################
\section{Conclusions \& discussion}
\label{conc}
%#################################################################

This work is based on using {\it Delphi}, a semi-analytic merger-tree based model, that jointly tracks DM and baryonic assembly of high-$z$ ($z \gsim 4-20$) galaxies to gain insights on the number density of DCBHs in the early Universe in three different DM cosmologies: CDM, 3.5 keV and 1.5 keV WDM. We note that {\it Delphi} has already been shown to match all observable data-sets, including the evolving ultra-violet luminosity and stellar mass functions, the stellar mass and ultra-violet luminosity densities and the mass-to-light ratios, for $z \gsim 5$ galaxies for both cold and warm dark matter models \citep{dayal2014, dayal2015, dayal2017}. In each DM model, we identify pristine halos with $T_{vir}\gsim10^4$ K irradiated by a critical LW background (where $J_{crit} = \alpha J_{21}$) as potential DCBH hosts. We explore $\alpha$ values between 30 and 300 to quantify the dependence on the critical LW background. We use the LW luminosity of each galaxy obtained from {\it Delphi} to calculate the LW background seen by any galaxy, including the enhancement due to galaxy clustering. We find that the probability of a halo being irradiated by a chosen $J_{\rm LW}$ value only increases by a factor of three as the halo mass increases from $10^{8-11.5}\Msun$ at a given $z$. However varying $\alpha$ by one order of magnitude, from 30 to 300, has an enormous impact, decreasing the probability of halos being irradiated by $J_{crit}$ by about four orders of magnitude, independent of $z$ and the underlying DM model; this naturally results in the number density of DCBH hosts decreasing by the same fraction. Interestingly we find that, for a given $\alpha$, the number density of possible DCBH hosts is comparable (to within an order of magnitude) in all three DM models for $z \simeq 8-10$. However, the delayed and accelerated structure formation in the 1.5 keV WDM model results, by extension, in a delayed metal-enrichment and a faster build-up of the LW background. These two effects conspire to create conditions such that DCBH hosts can persist down to redshifts as low as $z \simeq 5$ in the 1.5 keV WDM model, as compared to the end-redshift of $z \simeq 8$ in CDM. Generating DCBH spectra that are convolved with IGM transmission, we also present a detection strategy to hunt for such DCBHs candidates using three JWST NIRCam filters (F070W, F090W F115W): while $z \simeq 5$ DCBHs should be detectable in all three filters with the sensitivity threshold for a 10 ks JWST observation, $z \gsim 7$ DCBHs should be rendered undetectable in the F070W filter due to IGM attenuation of flux below the Ly$\alpha$. However, we note that follow-up spectroscopy will be required to confirm the true nature of such candidates. 

We briefly touch upon some caveats of this work: firstly, our free parameters, that are tuned to reproduce the Ultra-violet Luminosity Function, yield galaxy properties matching both global observables (e.g. the stellar mass density and UV luminosity density) and local observables (e.g. mass to light ratios and the stellar mass-halo mass relation) for high-$z$ galaxies. Further, varying these parameters by about 10\% has no sensible impact on galaxy properties, including their spectra/masses (Nobels et al., in prep.). These parameters could have an impact on the abundance of DCBHs only if they would lead to a variation in either $P_{leaf}$ or $P_{thres}$ by a factor of 10 (since a variation of a factor of 10 could be absorbed by the LW background range explored in this work). The largest degeneracy between cold and warm dark matter actually arises when ``maximal" reionization feedback scenarios are considered - i.e. where all galaxies below a certain halo mass or circular velocity are assumed to lose all of their gas mass due to photo-evaporation by the Ultra-violet (UV) background created during reionization (Bremer et al., in prep.). In this case, the lack of low mass star forming halos in CDM leads to its results tending towards WDM scenarios which we aim to explore in future works. However, realistically modelling the impact of UV feedback on the fraction of galaxies embedded in ionized regions remains an open problem in Astrophysics, and much beyond the scope of this work. Secondly, our model does not account for sub-clumps inside halos. However, this has no noticeable bearing on our results given that: (i) such clumps would be expected to have extremely low star formation capabilities, specially in the presence of a reionization UV background; and (ii) while the sub-clump density would be expected to be the highest for the most massive halos, DCBHs are typically hosted by the lowest mass pristine halos as per our results. Any small variation in the number density of DCBHs due to our excluding sub-clumps could therefore be absorbed by the large $\alpha$ range explored.

It is encouraging to note that our number densities only differ from previous works \citep{agarwal2012, dijkstra2014, habouzit2016} by less than one order of magnitude for a chosen $J_{crit}$ value. This difference could arise from a number of assumptions made in this work including: only assuming the starting leaves of a given halo to be pristine and using simple analytic estimates to account for a clustering-induced enhancement of the LW background seen  by a galaxy. We await the forthcoming JWST that, through detections of, both galaxies and, possible DCBHs in the early Universe, will yield invaluable hints on the particle-physics nature of dark matter.

%We end with noting the main caveats in this work: firstly, given that the metal-enrichment radius from galaxies is at most twice the virial radius, we have ignored external metal-enrichment in this work. Secondly, as already noted, our calculations of the LW background seen by a source do not properly account for clustering of halos. However, both of these simplistic estimates are justified, and compensated for, by studying $\alpha=30-300$, that has the most major impact on the number density of DCBH hosts. 

% ***************************************************************************
\section*{Acknowledgments} 
% ****************************************************************************
PD acknowledges support from the European Research Council's starter grant ERC StG-717001 and from the European Commission's and University of Groningen's CO-FUND Rosalind Franklin program.
FP acknowledges the Chandra grant nr. AR6-17017B and the NASA-ADAP grant nr. MA160009. VB acknowledges support from NSF grant AST-1413501.

% **************************************************************************

%%%%%%%%%%%%%%%%%%%%%%%%%%%%%%
\bibliographystyle{mn2e}
\bibliography{dcbh}

\label{lastpage} 
\end{document}